\documentclass[]{article}
\PassOptionsToPackage{numbers,compress}{natbib}
\usepackage[final]{nips_2016}
\usepackage{url}
\usepackage{amsmath}
\usepackage{amssymb}
\usepackage{amsbsy}
\usepackage{siunitx}
\usepackage{algorithm}
\usepackage{algorithmicx}
\usepackage{algpseudocode}
\usepackage{mathtools}
\usepackage{array}
\usepackage{setspace}
\usepackage{caption}
\usepackage{subcaption}
\usepackage{booktabs}
\usepackage{adjustbox}
\usepackage{microtype}

\usepackage{times}
\usepackage[bottom]{footmisc}
\usepackage{listings}
\usepackage{color}
\usepackage{xcolor}
\usepackage{textcomp}
\usepackage{xspace}

\usepackage{amsthm}

\usepackage{natbib}
\usepackage{todonotes}
\newcommand{\norm}[1]{\left\lVert#1\right\rVert}

\newcommand{\asto}{\overset{a.s.}{\to}}

\theoremstyle{plain}
\newtheorem{theorem}{Theorem}

\theoremstyle{remark}
\newtheorem{remark}{Remark}

\theoremstyle{lemma}

\newcommand{\hg}{\hat{\gamma}}

\newcommand{\real}{\mathbb{R}}

\DeclareMathOperator*{\var}{Var}
\newcommand{\E}{\mathbb{E}}
\newcommand{\N}{\mathbb{N}}

\graphicspath{{./code/}}

\title{On the Pitfalls of Nested Monte Carlo}

\author{{\bf Tom Rainforth, ~Robert Cornish, ~Hongseok Yang, ~Frank Wood}\\
	University of Oxford \\
	{\small \texttt{\{twgr,rcornish,fwood\}@robots.ox.ac.uk, hongseok.yang@cs.ox.ac.uk}}\\
}

\begin{document}

\maketitle

\begin{abstract}
  There is an increasing interest in estimating expectations outside of the classical
  inference framework, such as for models expressed as probabilistic programs. Many of
  these contexts call for some form of \emph{nested inference} to be applied.  In this
  paper, we analyse the behaviour of nested Monte Carlo (NMC) schemes, for which classical
  convergence proofs are insufficient. We give conditions under which NMC will converge,
  establish a rate of convergence, and provide empirical data that suggests that this rate
  is observable in practice. Finally, we prove that general-purpose nested inference
  schemes are inherently biased.  Our results serve to warn of the dangers associated with
  na\"ive composition of inference and models.
\end{abstract}

\section{Introduction}
\label{sec:intro}

Monte Carlo (MC) methods have become a ubiquitous means of carrying out approximate Bayesian inference.  
From simplistic Metropolis Hastings approaches to state-of-the-art algorithms such as the bouncy particle sampler \cite{bouchard2015bouncy} and interacting particle Markov chain MC \cite{rainforth2016interacting},
the aim of these methods is always the same: to generate approximate samples from a posterior, from which an expectation can be calculated.  
Although interesting alternatives have recently been suggested \cite{briol2015probabilistic}, MC integration is used almost exclusively for calculating these expectations from the produced samples.

The convergence of conventional MC integration has been covered extensively in previous
literature \cite{gilks2005markov,huggins2015convergence}, but the theoretical implications
arising from the \emph{nesting} of MC schemes, where terms in the integrand depend on the
result of a separate, nested, MC integration, have been predominantly overlooked.
This paper examines convergence of such nested Monte Carlo (NMC) methods.  Although we
demonstrate that the construction of consistent NMC algorithms is possible, we reveal a
number of associated pitfalls. In particular, NMC estimators are inherently biased, may
require additional assumptions for convergence, and have significantly diminished
convergence rates.

A significant motivating application for NMC occurs in the context of probabilistic
programming systems (PPS)
\cite{goodman2008church,mansinghka2014venture,paige2014compilation,pfeffer2009figaro,rainforth2016nips,wood2014new},
which allow a decoupling of model specification, in the form of a generative model with
conditioning primitives, and inference, in the form of a back-end engine
capable of operating on arbitrary programs. Many PPS allow for arbitrary nesting of
programs so that it is easy to define and run nested inference problems, which has already
begun to be exploited in application specific work \cite{ouyang2016practical}. However, such
nesting can violate assumptions made in asserting the correctness of the underlying
inference schemes. Our work highlights this issue and gives theoretical insight into the
behaviour of such systems.
This serves as a warning against na\"{i}ve composition and a guide as to when we can
expect to make reasonable estimations.

Some nested inference problems can be tackled by so-called pseudo-marginal methods
\cite{andrieu2009pseudo,andrieu2015convergence,doucet2015efficient,lindsten2016pseudo}.  These
consider cases of Bayesian inference where the likelihood is intractable, such as
when it originates from an Approximate Bayesian Computation (ABC)
\cite{beaumont2002approximate,csillery2010approximate}.
They proceed by reformulating the problem in an extended space
\cite{rainforth2016interacting}, with auxiliary variables representing the stochasticity in
the likelihood computation, allowing the problem to be expressed as a single expectation.

Our work goes beyond this by considering cases in which a
non-linear mapping is applied to the output of the inner expectation, so that
this reformulation to a single expectation is no longer possible.  
One scenario where this occurs is the expected information gain used in Bayesian
experimental design.  This requires the calculation of an entropy of a marginal
distribution, and therefore includes the expectation of the logarithm of an expectation.
Presuming these expectations cannot be calculated exactly, one
must therefore resort to some sort of approximate nested inference scheme.

%




\section{Problem Formulation}
\label{sec:prob-form}

The idea of MC is that the expectation of an arbitrary function 
$\lambda \colon \mathcal{Y} \rightarrow \mathcal{F} \subseteq \real^{D_{f}}$ under a probability distribution $p(y)$ for its input $y \in \mathcal{Y}$ can be approximately calculated in the following fashion:
\begin{align}
\label{eq:MC}
I &= \mathbb{E}_{y\sim p(y)} \left[\lambda(y)\right] \\
&\approx \frac{1}{N} \sum_{n=1}^{N} \lambda(y_n) \quad \mathrm{where} \quad y_n \sim p(y).
\end{align}
In this paper, we consider the case where $\lambda$ is itself intractable, defined only in terms of a functional mapping of an expectation. Specifically,
\begin{align}
\label{eq:f}
\lambda(y) = f(y,\gamma(y))
\end{align}
where we can evaluate $f \colon \mathcal{Y} \times \Phi \rightarrow \mathcal{F}$ exactly for a given $y$ and $\gamma (y)$, but where $\gamma (y)$ is the output of an intractable expectation of another variable $z \in \mathcal{Z}$, that is, 
\begin{subequations}
\label{eq:gamma}
\begin{align}
\label{eq:gamma_1}
\mathrm{either}\quad
\gamma(y) &=  \mathbb{E}_{z\sim p(z | y)} \left[\phi(y,z)\right] \\
\label{eq:gamma_2}
\mathrm{or} \quad \gamma(y) &= \mathbb{E}_{z\sim p(z)} \left[\phi(y,z)\right]
\end{align}
\end{subequations}
depending on the problem, with $\phi \colon \mathcal{Y} \times \mathcal{Z} \rightarrow \Phi \subseteq \real^{D_{\phi}}$.
All our results apply to both cases, but we will focus on the former for clarity.
Estimating $I$ requires a nested integration.  We refer to tackling both required integrations using Monte Carlo as nested Monte Carlo:
\begin{subequations}
	\label{eq:nested-mc}
\begin{align}
\label{eq:nested-outer}
I \approx I_{N,M} =& \frac{1}{N}  \sum_{n=1}^{N} f(y_n,(\hat{\gamma}_M)_n) \quad \mathrm{where} \quad  y_n \sim p(y) \quad \mathrm{and} \\
\label{eq:nested-inner}
(\hat{\gamma}_M)_n =& \frac{1}{M}  \sum_{m=1}^{M}  \phi(y_n,z_{n,m}) \quad \mathrm{where} \quad z_{n,m} \sim p(z | y_n).
\end{align}
\end{subequations}

The rest of this paper proceeds as follows. In Section~\ref{sec:linear_case}, we consider
a special case of $f$ that allows us to recover the standard Monte Carlo convergence rate.
In Section~\ref{sec:convergence}, we establish convergence results for $I_{N,M}$ given a
general class of $f$. In Section~\ref{sec:bias}, we show that any general-purpose NMC
scheme must be biased. Finally, in Section~\ref{sec:empirical}, we present empirical results
suggesting that our theoretical convergence rates are observed in practise.


\section{Reformulation to a Single Expectation}
\label{sec:linear_case}

Suppose that $f$ 
is integrable and linear in its second argument, i.e. $f(y,\alpha v + \beta w) = 
\alpha f(y,v)+ \beta f(y,w)$ (or equivalently $f(y,z) = g(y)z$ for some $g(y)$).
In this case, we can rearrange the problem to a single expectation:
\begin{align*}
I
 = \mathbb{E}_{y \sim p(y)}\left[f(y,\gamma(y))\right]
& = \mathbb{E}_{y \sim p(y)}\left[f\left(y,\mathbb{E}_{z\sim p(z|y)}\left[\phi(y,z)\right]\right)\right]
\\
& = \mathbb{E}_{y \sim p(y)}\left[ \mathbb{E}_{z\sim p(z|y)}\left[f(y,\phi(y,z))\right]\right]
\\
& \approx\frac{1}{N} \sum_{n=1}^{N} f(y_n,\phi(y_n,z_n)) \quad \mathrm{where} \quad (y_n, z_n) \sim p(y)p(z|y).
\end{align*}
This will give the MC convergence rate of $O(1/N)$ in the mean square error of the estimator, provided we can generate the required samples.
Many models do permit this rearrangement, such as those considered by pseudo-marginal
methods.
Note that if $\gamma(y)$ is of the form of~\eqref{eq:gamma_2} instead of~\eqref{eq:gamma_1}, then $y$ and $z$ are drawn independently from their marginal distributions instead of the joint.


\section{Convergence of Nested Monte Carlo}
\label{sec:convergence}

Since we cannot always unravel our problem as in the previous section, we must resort to
NMC in order to compute $I$ in general. Our aim here is to show that
approximating $I \approx I_{N,M}$ is in principle possible, at least when $f$ is
well-behaved. In particular, we prove a form of almost sure convergence of $I_{N,M}$ to
$I$ and establish an upper bound on the convergence rate of its mean squared error.


To more formally characterize our conditions on $f$, consider sampling a single $y_1$. Then $(\hat{\gamma}_M)_1=\frac{1}{M}\sum_{m=1}^{M}
\phi(y_1,z_{1,m})\rightarrow\gamma(y_1)$ as $M \rightarrow \infty$, as the left-hand side is
a Monte Carlo estimator. If $f$ is continuous around $y_1$, this also implies
$f(y_1,(\hat{\gamma}_M)_1) \rightarrow f(y_1,\gamma(y_1))$. 
Informally, our requirement is that this holds in expectation, i.e. that it holds
when we incorporate the effect of the outer estimator.
More precisely,
we define $(\epsilon_M)_n = \left|f(y_n, (\hat{\gamma}_M)_n) -
f(y_n,\gamma(y_n))\right|$, and require that $\E\left[(\epsilon_M)_1\right] \to 0$ as $M \to
\infty$ (noting that as $(\epsilon_M)_n$ are i.i.d. $\E\left[(\epsilon_M)_1\right] =
\E\left[(\epsilon_M)_n\right], \forall n\in\N$). Informally, this ``expected continuity''
assumption is weaker than uniform continuity as it does allow discontinuities in $f$, though we 
leave full characterization of intuitive criteria for $f$ to future work. 
We are now ready to state our theorem for almost sure convergence. Proofs for all theorems are provided in the Appendices. 

\begin{theorem} \label{the:Consistent}
  For $n \in \N$, let $(\epsilon_M)_n = \left|f(y_n, (\hat{\gamma}_M)_n) - f(y_n,
  \gamma(y_n))\right|$. 
  If~~$\E\left[(\epsilon_M)_1\right] \to 0$ as $M \to \infty$, then
  there exists a $\tau : \N \to \N$ such that $I_{\tau(M),M} \asto I$ as $M \to \infty$.
\end{theorem}
\begin{remark}
As this convergence is in $M$, it implies (and is reinforced by the
convergence rate given below) that it is necessary for the number of samples in the inner estimator 
to increase with the number of samples in the outer estimator to ensure convergence for most $f$.
Theorem~\ref{the:approx-inf} gives an intuitive reason for why this should be the case by noting that for finite $M$, the
bias on each inner term will remain non-zero as $N\rightarrow\infty$.
\end{remark}
%
%
%
\begin{theorem} \label{the:Rate}
	If $f$ is Lipschitz continuous and $f(y_n, \gamma(y_n)), \phi(y_n, z_{n,m}) \in
	L^2$, then the mean squared error of $I_{N,M}$ converges at rate $O\left(1/N +
	1/M\right)$.
\end{theorem}
\vspace{-5pt}
Inspection of the convergence rate above shows that, given a total number of samples
$T=MN$, our bound is tightest when $\tau(M) = O(M)$ (see Section~\ref{sec:opt-conv}), with a
corresponding rate $O(1/\sqrt{T})$. Although Theorem~\ref{the:Consistent} does not
guarantee that \emph{this} choice of $\tau$ converges almost surely, any other choice of
$\tau$ will give a a weaker guarantee than this already problematically slow rate. 
Future work might consider specific forms of  $\tau$ that ensure
convergence.

With repeated nesting, informal extension of Theorem~\ref{the:Rate} suggests that the
convergence rate will become $O(\sum_{i=1}^{d} \frac{1}{N_i})$ where $N_i$ is the number
of samples used for the estimation at nesting depth $i$. This yields a bound on our
convergence rate in total number of samples that becomes exponentially weaker as the total 
nesting depth $d$ increases. We leave a formal proof of this to future work.


\section{The Inherent Bias of Nested Inference} \label{sec:bias}

The previous section confirmed the capabilities of NMC; in this section we
establish a limitation by showing that any such general-purpose nesting
scheme must be biased in the following sense:
\begin{theorem}
	\label{the:approx-inf}
        Assume that $\gamma(y) = \mathbb{E}_{z\sim p(z|y)}[\phi(y,z)]$ is 
        integrable as a function of $y$ but cannot be calculated exactly. Then, there does
        not exist a pair $(\mathcal{I}, \mathcal{J})$ of inner and outer estimators such
        that 
\begin{enumerate}
		\vspace{-5pt}
        \item the inner estimator $\mathcal{I}$ provides estimates $\hg_y \in \Phi$ at a given $y \in \mathcal{Y}$;
        		\vspace{-5pt}
        \item given an integrable $f \colon \mathcal{Y} \times \Phi \rightarrow \mathcal{F} \subseteq \real^{D_{f}}$ the outer estimator $\mathcal{J}$ maps a set of samples $\hat{\zeta} = \left\{(y_1,\hg_{y_1}),\dots,(y_n,\hg_{y_n})\right\}$, with $\hg_{y_i}$ generated using $\mathcal{I}$, to an unbiased estimate $\psi(\hat{\zeta},f)$ of $I(f)$, i.e. $\mathbb{E}[\psi (\hat{\zeta},f)] = I(f)$; 
        		\vspace{-5pt}
        \item $\mathbb{E}_{y \sim p(y)}[\mathbb{E}[f(y,\hg_y) | y]] - \mathbb{E}[\psi(\hat{\zeta},f)]
                \geq 0$ for all integrable $f$, i.e. if $\mathcal{I}$ is combined with an exact outer estimator, there is no $f$ for which the resulting estimator is negatively biased (see Remark~\ref{rem:neg-bias-exp}).
                		\vspace{-5pt}
\end{enumerate}
        This result remains even if the inequality in the third condition is reversed from $ge0$ to $le0$.
\end{theorem}

\begin{remark}
	\label{rem:neg-bias-exp}
Informally, the first two conditions here simply provide definitions for $\mathcal{I}$ and
$\mathcal{J}$ and state that they provide unbiased estimation of $I$ for all $f$.  The
third condition is somewhat more subtle.  A simpler, but less general, alternative condition would
have been to state that $\mathcal{I}$ provides unbiased estimates for any $f$, i.e. $\mathbb{E}[f(y,\hg_y) | y] = f(y,\gamma(y)), \; \forall f$.  The
additional generality provided by the used formulation eliminates most cases in which both
$\mathcal{I}$ and $\mathcal{J}$ are biased, but in such a way that these biases cancel
out.
Specifically, we allow $\mathcal{I}$ to be biased so long as this bias has the
same sign for all $f$.  
As $\mathcal{I}$ and $\mathcal{J}$ are independent processes, it
is intuitively reasonable to assume that $\mathcal{J}$ does not eliminate bias from
$\mathcal{I}$ in a manner that is specific to $f$ and so we expect this condition to hold in practice. Future work might consider a completely general proof
that also considers this case.
\end{remark}

This result suggests that general purpose, unbiased inference is impossible for nested probabilistic program
queries which cannot be mathematically expressed as single inference of the
form~\eqref{eq:MC}.  Such rearrangement is not possible when the outer query depends
nonlinearly on a \emph{marginal} of the inner query.  Consequently, query nesting using existing systems\footnote{We note that for certain nonlinear $f$, it may still be possible to develop an unbiased inference scheme using a combination of a convergent Maclaurin expansion and Russian Roulette sampling \cite{lyne2015russian}.} cannot
provide unbiased estimation of problems that cannot be expressed as a single query.
However, the additional models that it does allow expression for, such
as the experimental design example, might still be estimable in consistent fashion as shown in the previous section.


\section{Empirical Verification}
\label{sec:empirical}

\begin{figure}[t]
	\vspace{-10pt}
	\centering
	\includegraphics[width=0.55\textwidth]{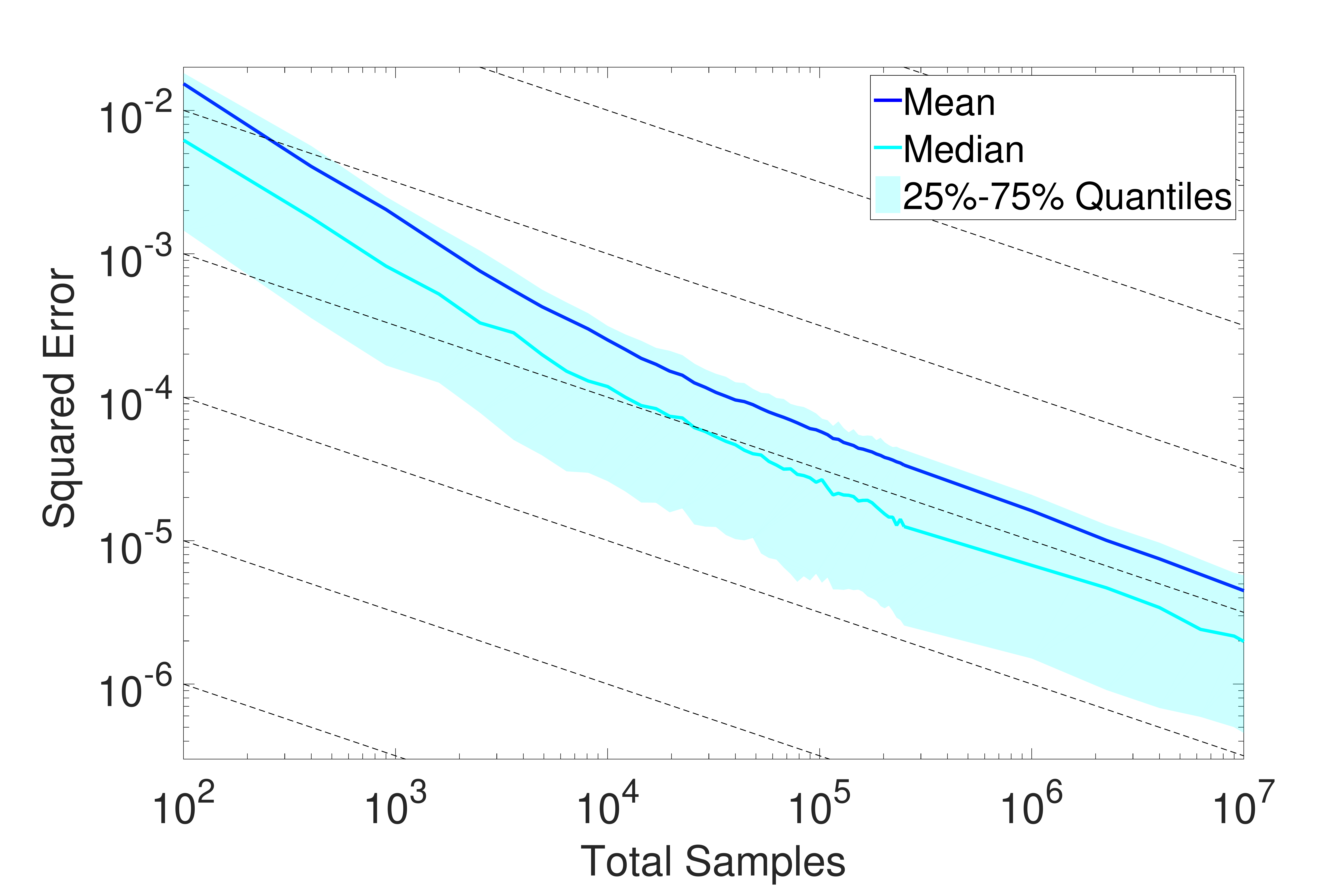}
  \caption{Empirical convergence of NMC on~\eqref{eq:model}.  Results are obtained by
  averaging over 1000 independent runs with $\tau(N) = N$ such that convergence rate
  obtained from Theorem~\ref{the:Rate} is $O(1/\sqrt{T})$.\label{fig:emprical-conv}}
\end{figure}

Strictly speaking, the convergence rates proven in Section~\ref{sec:convergence} are only
\emph{upper bounds} on the worst case performance we can expect. We therefore provide a 
short empirical verification to see whether these convergence rates are
tight in practice. For this, we consider the following simple model whose exact solution
can be calculated:
\begin{subequations}
\label{eq:model}
\begin{align}
y & \sim \mathrm{Uniform}(-1,1) \\
z &\sim \mathcal{N}(0,1) \\
\phi(z,y) &= \sqrt{2/\pi}\exp\left(-2(y-z)^2\right) \\
f(y,\gamma(y)) &= \log (\gamma (y)).
\end{align}
\end{subequations}
Figure~\ref{fig:emprical-conv} shows the corresponding empirical convergence obtained by
applying~\eqref{eq:nested-mc} to~\eqref{eq:model} directly, and shows that, at least in
this case, the theoretical convergence rates from Theorem~\ref{the:Rate} are indeed
realised.

%


\section{Conclusions}

We have shown that it is theoretically possible for a nested Monte Carlo scheme to yield a
consistent estimator, and have quantified the convergence error associated with doing so.
However, we have also revealed a number of pitfalls that can arise if nesting is applied
na\"{i}vely, such as that the resulting estimator is necessarily biased, requires additional
assumptions on $f$, is unlikely to converge unless the number of samples used in the inner
estimator is driven to infinity,
and is likely to converge at a significantly slower rate than un-nested Monte
Carlo. These results have implications for applications ranging from experimental design
to probabilistic programming, and serve both as an invitation for further inquiry and a
caveat against careless use.

	
\section*{Acknowledgements}

Tom Rainforth is supported by a BP industrial grant. Robert Cornish is supported by an NVIDIA scholarship. Frank Wood is supported under DARPA PPAML through the U.S. AFRL under Cooperative Agreement FA8750-14-2-0006, Sub Award number 61160290-111668.

\bibliography{refs_nested_mc}

\begin{thebibliography}{19}
\providecommand{\natexlab}[1]{#1}
\providecommand{\url}[1]{\texttt{#1}}
\expandafter\ifx\csname urlstyle\endcsname\relax
  \providecommand{\doi}[1]{doi: #1}\else
  \providecommand{\doi}{doi: \begingroup \urlstyle{rm}\Url}\fi

\bibitem[Andrieu and Roberts(2009)]{andrieu2009pseudo}
C.~Andrieu and G.~O. Roberts.
\newblock The pseudo-marginal approach for efficient {M}onte {C}arlo
  computations.
\newblock \emph{The Annals of Statistics}, pages 697--725, 2009.

\bibitem[Andrieu et~al.(2015)Andrieu, Vihola, et~al.]{andrieu2015convergence}
C.~Andrieu, M.~Vihola, et~al.
\newblock Convergence properties of pseudo-marginal {M}arkov chain {M}onte
  {C}arlo algorithms.
\newblock \emph{The Annals of Applied Probability}, 25\penalty0 (2):\penalty0
  1030--1077, 2015.

\bibitem[Beaumont et~al.(2002)Beaumont, Zhang, and
  Balding]{beaumont2002approximate}
M.~A. Beaumont, W.~Zhang, and D.~J. Balding.
\newblock {Approximate Bayesian computation in population genetics}.
\newblock \emph{Genetics}, 162\penalty0 (4):\penalty0 2025--2035, 2002.

\bibitem[Bouchard-C{\^o}t{\'e} et~al.(2015)Bouchard-C{\^o}t{\'e}, Vollmer, and
  Doucet]{bouchard2015bouncy}
A.~Bouchard-C{\^o}t{\'e}, S.~J. Vollmer, and A.~Doucet.
\newblock {The Bouncy Particle Sampler: A Non-Reversible Rejection-Free Markov
  Chain Monte Carlo Method}.
\newblock \emph{arXiv preprint arXiv:1510.02451}, 2015.

\bibitem[Briol et~al.(2015)Briol, Oates, Girolami, Osborne, and
  Sejdinovic]{briol2015probabilistic}
F.-X. Briol, C.~J. Oates, M.~Girolami, M.~A. Osborne, and D.~Sejdinovic.
\newblock Probabilistic integration: A role for statisticians in numerical
  analysis?
\newblock \emph{arXiv preprint arXiv:1512.00933}, 2015.

\bibitem[Csill{\'e}ry et~al.(2010)Csill{\'e}ry, Blum, Gaggiotti, and
  Fran{\c{c}}ois]{csillery2010approximate}
K.~Csill{\'e}ry, M.~G. Blum, O.~E. Gaggiotti, and O.~Fran{\c{c}}ois.
\newblock {Approximate Bayesian computation (ABC) in practice}.
\newblock \emph{Trends in ecology \& evolution}, 25\penalty0 (7):\penalty0
  410--418, 2010.

\bibitem[Doucet et~al.(2015)Doucet, Pitt, Deligiannidis, and
  Kohn]{doucet2015efficient}
A.~Doucet, M.~Pitt, G.~Deligiannidis, and R.~Kohn.
\newblock Efficient implementation of {M}arkov chain {M}onte {C}arlo when using
  an unbiased likelihood estimator.
\newblock \emph{Biometrika}, page asu075, 2015.

\bibitem[Gilks(2005)]{gilks2005markov}
W.~R. Gilks.
\newblock \emph{{Markov chain Monte Carlo}}.
\newblock Wiley Online Library, 2005.

\bibitem[Goodman et~al.(2008)Goodman, Mansinghka, Roy, Bonawitz, and
  Tenenbaum]{goodman2008church}
N.~D. Goodman, V.~K. Mansinghka, D.~Roy, K.~Bonawitz, and J.~B. Tenenbaum.
\newblock Church: a language for generative models.
\newblock 2008.

\bibitem[Huggins and Roy(2015)]{huggins2015convergence}
J.~H. Huggins and D.~M. Roy.
\newblock {Convergence of Sequential Monte Carlo-based Sampling Methods}.
\newblock \emph{arXiv preprint arXiv:1503.00966}, 2015.

\bibitem[Lindsten and Doucet(2016)]{lindsten2016pseudo}
F.~Lindsten and A.~Doucet.
\newblock Pseudo-marginal {H}amiltonian {M}onte {C}arlo.
\newblock \emph{arXiv preprint arXiv:1607.02516}, 2016.

\bibitem[Lyne et~al.(2015)Lyne, Girolami, Atchade, Strathmann, Simpson,
  et~al.]{lyne2015russian}
A.-M. Lyne, M.~Girolami, Y.~Atchade, H.~Strathmann, D.~Simpson, et~al.
\newblock {On Russian roulette estimates for Bayesian inference with
  doubly-intractable likelihoods}.
\newblock \emph{Statistical science}, 30\penalty0 (4):\penalty0 443--467, 2015.

\bibitem[Mansinghka et~al.(2014)Mansinghka, Selsam, and
  Perov]{mansinghka2014venture}
V.~Mansinghka, D.~Selsam, and Y.~Perov.
\newblock Venture: a higher-order probabilistic programming platform with
  programmable inference.
\newblock \emph{arXiv preprint arXiv:1404.0099}, 2014.

\bibitem[Ouyang et~al.(2016)Ouyang, Tessler, Ly, and
  Goodman]{ouyang2016practical}
L.~Ouyang, M.~H. Tessler, D.~Ly, and N.~Goodman.
\newblock Practical optimal experiment design with probabilistic programs.
\newblock \emph{arXiv preprint arXiv:1608.05046}, 2016.

\bibitem[Paige and Wood(2014)]{paige2014compilation}
B.~Paige and F.~Wood.
\newblock A compilation target for probabilistic programming languages.
\newblock \emph{arXiv preprint arXiv:1403.0504}, 2014.

\bibitem[Pfeffer(2009)]{pfeffer2009figaro}
A.~Pfeffer.
\newblock Figaro: An object-oriented probabilistic programming language.
\newblock \emph{Charles River Analytics Technical Report}, 137, 2009.

\bibitem[Rainforth et~al.(2016{\natexlab{a}})Rainforth, Le, van~de Meent,
  Osborne, and Wood]{rainforth2016nips}
T.~Rainforth, T.~A. Le, J.-W. van~de Meent, M.~A. Osborne, and F.~Wood.
\newblock {B}ayesian {O}ptimization for {P}robabilistic {P}rograms.
\newblock In \emph{Advances in Neural Information Processing Systems},
  2016{\natexlab{a}}.

\bibitem[Rainforth et~al.(2016{\natexlab{b}})Rainforth, Naesseth, Lindsten,
  Paige, van~de Meent, Doucet, and Wood]{rainforth2016interacting}
T.~Rainforth, C.~A. Naesseth, F.~Lindsten, B.~Paige, J.-W. van~de Meent,
  A.~Doucet, and F.~Wood.
\newblock Interacting particle {M}arkov chain {M}onte {C}arlo.
\newblock In \emph{Proceedings of the 33rd International Conference on Machine
  Learning}, volume~48. JMLR: W$\backslash$\&CP, 2016{\natexlab{b}}.

\bibitem[Wood et~al.(2014)Wood, van~de Meent, and Mansinghka]{wood2014new}
F.~Wood, J.~W. van~de Meent, and V.~Mansinghka.
\newblock A new approach to probabilistic programming inference.
\newblock In \emph{AISTATS}, pages 2--46, 2014.

\end{thebibliography}

\newpage
\appendix

\section{Proof of Almost Sure Convergence (Theorem~\ref{the:Consistent})}
\label{sec:app:conv-proof}

\begin{proof}
For all $N, M$, we have by the triangle inequality that
\[
\left|I_{N,M} - I\right| \leq V_{N,M} + U_N,
\]
where
\begin{eqnarray*}
  V_{N,M} &=& \left|\frac{1}{N} \sum_{n=1}^N f(y_n, \gamma(y_n)) - I_{N,M} \right| \\
  U_N &=& \left|I - \frac{1}{N} \sum_{n=1}^N f(y_n, \gamma(y_n)) \right|.
\end{eqnarray*}
A second application of the triangle inequality then allows us to write
\begin{equation} \label{eq:vnm}
  V_{N,M} \leq \frac{1}{N} \sum_{n=1}^N (\epsilon_M)_n
\end{equation}
where we recall that $(\epsilon_M)_n = |f(y_n, \gamma(y_n)) - f(y_n, \hat{\gamma}_n)|$.
Now, for all fixed $M$, each $(\epsilon_M)_n$ is i.i.d, and our assumption that $\E
\left[(\epsilon_M)_1\right] \to 0$ as $M \to \infty$ ensures $\E
\left[\left|(\epsilon_M)_n \right|\right] < \infty$ for all $M$ sufficiently large.
Consequently, the strong law of large numbers means that
\[
  \frac{1}{N} \sum_{n=1}^N (\epsilon_M)_n \asto \E\left[(\epsilon_M)_1\right]
\]
as $N \to \infty$. This allows us to define $\tau_1 : \N \to \N$ by choosing $\tau_1(M)$
to be large enough that
\[
  \left|\frac{1}{\tau_1(M)} \sum_{n=1}^{\tau_1(M)} (\epsilon_M)_n - \E\left[(\epsilon_M)_1\right]\right| < \frac{1}{M}
\]
almost surely, for each $M \in \N$. Consequently,
\[
  \frac{1}{\tau_1(M)} \sum_{n=1}^{\tau_1(M)} (\epsilon_M)_n < \frac{1}{M} + \E\left[(\epsilon_M)_1\right]
\]
almost surely and therefore
\[
  V_{\tau_1(M),M} < \frac{1}{M} + \E \left[(\epsilon_M)_1\right]
\]
almost surely.

To complete the proof, we must remove the dependence of $U_N$ on $N$ also. This is
straightforward once we observe that $U_N \asto 0$ as $N \to \infty$ by the strong law of
large numbers, which allows us to define $\tau_2 : \N \to \N$ by taking $\tau_2(M)$ large
enough that
\[
  U_{\tau_2(M)} < \frac{1}{M}
\]
almost surely, for each $M \in \N$.

We can now define $\tau(M) = \max(\tau_1(M), \tau_2(M))$. It then follows that, for
all $M$,
\[
  \left|I - I_{\tau(M),M}\right| \leq \frac{1}{M} + \frac{1}{M} + \E\left[(\epsilon_M)_1\right]
\]
almost surely. By assumption we have $\E\left[(\epsilon_M)_1\right] \to 0$, so that
$I_{\tau(M),M} \asto I$ as desired.

\end{proof}

\section{Proof of Convergence Rate (Theorem~\ref{the:Rate})}
\label{sec:app:rate-proof}

\begin{proof}

Using Minkowski's inequality, we can bound the mean squared error of $I_{N,M}$ by
\begin{equation} \label{eq:mse-bound}
  \E[(I-I_{N,M})^2] = \norm{I - I_{N,M}}_2^2 \leq {U}^2 + {V}^2 + 2 U V \leq 2\left(U^2 + V^2\right)
\end{equation}
where
\begin{eqnarray*}
  U &=& \norm{I - \frac{1}{N} \sum_{n=1}^N f(y_n, \gamma(y_n))}_2 \\
  V &=& \norm{\frac{1}{N} \sum_{n=1}^N f(y_n, \gamma(y_n)) - I_{N,M}}_2.
\end{eqnarray*}
We see immediately that $U = O\left(1 / \sqrt{N}\right)$, since $\frac{1}{N} \sum_{n=1}^N
f(y_n, \gamma(y_n))$ is a Monte Carlo estimator for $I$, noting our assumption that
$f(y_n, \gamma(y_n)) \in L^2$. For the second term,
\begin{eqnarray*}
  V &=& \norm{\frac{1}{N} \sum_{n=1}^N f(y_n, (\hat{\gamma}_M)_n) - f(y_n, \gamma(y_n))}_2 \\
  &\leq& \frac{1}{N} \sum_{n=1}^{N} \norm{f(y_n, (\hat{\gamma}_M)_n) - f(y_n,
    \gamma(y_n))}_2 \\
  &\leq& \frac{1}{N} \sum_{n=1}^N K \norm{(\hat{\gamma}_M)_n - \gamma(y_n)}_2 \\
\end{eqnarray*}
where $K$ is a fixed constant, again by Minkowski and using the assumption that $f$ is
Lipschitz. We can rewrite

\[
  \norm{(\hat{\gamma}_M)_n - \gamma(y_n)}_2^2
    = \E \left[ \E \left[ ((\hat{\gamma}_M)_n - \gamma(y_n))^2 \middle| y_n \right]\right].
\]

by the tower property of conditional expectation, and note that
\begin{eqnarray*}
  \E \left[ ((\hat{\gamma}_M)_n - \gamma(y_n))^2 \middle| y_n \right]
    &=& \var\left(\frac{1}{M} \sum_{m=1}^M \phi(y_n, z_{n,m}) \middle| y_n \right) \\
    &=& \frac{1}{M} \var\left(\phi(y_n, z_{n,1}) \middle| y_n \right)
\end{eqnarray*}
since each $z_{n,m}$ is i.i.d. and conditionally independent given $y_n$. As
such
\[
  \norm{(\hat{\gamma}_M)_n - \gamma(y_n)}_2^2
  = \frac{1}{M} \, \E \left[ \var \left(\phi(y_n, z_{n,1}) \middle| y_n \right)\right] = O(1/M),
\]
noting that $\E \left[ \var \left(\phi(y_n, z_{n,1}) \middle| y_n \right)\right]$ is a
finite constant by our assumption that $\phi(y_n, z_{n,m}) \in L^2$. Consequently,
\[
  V \leq \frac{NK}{N} O\left(1/\sqrt{M}\right) = O\left(1/\sqrt{M}\right).
\]
Substituting these bounds for $U$ and $V$ in \eqref{eq:mse-bound} gives
\[
  \norm{I - I_{N,M}}_2^2
    \leq 2\left(O\left(1/\sqrt{N}\right)^2 + O\left(1/\sqrt{M}\right)^2\right)
    = O\left(1/N + 1/M\right)
\]
as desired.
\end{proof}

\section{Optimising the Convergence Rate}
\label{sec:opt-conv}

We have shown that the mean squared error converges at a rate $O(1/N + 1/M)$. For a given
choice of $\tau$ as in Theorem~\ref{the:Consistent}, this becomes $O(R)$, where
\[
  R = 1/\tau(M) + 1/M.
\]
Now let
\[
  T = \tau(M) \cdot M
\]
denote the total number of samples used by our scheme. 
We wish to understand the relationship between $T$ and $R$.

First, suppose $\tau(M) = O(M)$ as $M \to \infty$. This easily gives
\[
  \frac{1}{\sqrt{\tau(M)}} = O\left(\frac{1}{\sqrt{M}}\right)
\]
as $M \to \infty$, so that
\[
  \frac{1}{\sqrt{T}} = \frac{1}{\sqrt{M}} \frac{1}{\sqrt{\tau(M)}}
    = O\left(\frac{1}{M}\right)
\]
and as such 
\begin{equation} \label{eq:rm}
  R = O\left(\frac{1}{\sqrt{T}}\right)
\end{equation}
as $M \to \infty$.

In contrast, consider the case that $M \ll \tau(M)$ as $M \to \infty$. We then have
\[
  \frac{1}{\sqrt{M}} \gg \frac{1}{\sqrt{\tau(M)}} 
\]
as $M \to \infty$, so that
\[
  R = O\left(\frac{1}{M}\right) \gg \frac{1}{\sqrt{M}} \frac{1}{\sqrt{\tau(M)}} = \frac{1}{\sqrt{T}}
\]
as $M \to \infty$. Comparing with \eqref{eq:rm}, we observe that, for the same total
budget of samples $T$, this choice of $\tau$ provides a strictly weaker convergence
guarantee than in the previous case. A similar argument shows that the same is true when
$M \gg \tau(M)$ also.

\section{Proof of Inherent Bias (Theorem~\ref{the:approx-inf})}
\label{sec:app:bias-proof}

\begin{proof}
For the sake of contradiction, suppose that a pair $(\mathcal{I}, \mathcal{J})$ of
inner and outer estimators satisfies the conditions in the theorem.
Consider the possible pair of instances for $f$, $f_1(y,w) = (\gamma(y)-w)^2$ and $f_2(y,w) = -f_1(y,w)$. Since $\gamma(y)$ cannot be computed exactly by assumption, 
$\hg_y$ as an estimate for $\gamma(y)$ has non-zero variance. Thus, for every $y \in \mathbb{R}$,
the following inequalities hold almost surely: 
\[
f_1(y,\hg_y) > f_1(y,\gamma(y)) = 0 = f_2(y,\gamma(y)) > f_2(y,\hg_y).
\]
This implies that 
\begin{equation}
\label{eqn:impossibility-thm1}
\mathbb{E}_{y \sim p(y)}\left[\mathbb{E}\left[f_1(y,\hg_y) | y\right]\right] 
> 0
> \mathbb{E}_{y \sim p(y)}\left[\mathbb{E}\left[f_2(y,\hg_y) | y\right]\right]. 
\end{equation}
But
\begin{equation}
\label{eqn:impossibility-thm2}
\mathbb{E}\left[\psi(\hat{\zeta},f_1)\right]
= I(f_1) = 0 = I(f_2) = 
\mathbb{E}\left[\psi(\hat{\zeta},f_2)\right].
\end{equation}
Thus, 
\[
\left(\mathbb{E}_{y \sim p(y)}\left[\mathbb{E}\left[f_1(y,\hg_y) | y\right]\right] 
-
\mathbb{E}\left[\psi(\hat{\zeta},f_1)\right]\right) > 0
> \left(\mathbb{E}_{y \sim p(y)}\left[\mathbb{E}\left[f_2(y,\hg_y) | y\right]\right] 
-
\mathbb{E}\left[\psi(\hat{\zeta},f_2)\right]\right). 
\]
This contradicts the third condition in the theorem regardless of whether we use the original
$\geq 0$ or the alternative $\leq 0$.
\end{proof}

\end{document}